\documentclass[12pt]{article} \usepackage{amssymb}

\newcommand{\citen}[1]{\cite{#1}}

\makeatletter
\@addtoreset{equation}{section}

\makeatother

\newcommand{\PTP}[3]{Prog.~Theor.~Phys. {\bf #1} (#2), #3}
\newcommand{\JL}[4]{#1 {\bf #2} (#3), #4}
\newcommand{\andvol}[3]{{\bf #1} (#2), #3}

\title{Schwarzschild Space-Time in Gauge Theories of Gravity}
\author{Toshiharu \textsc{Kawai},\thanks{e-mail:
    \texttt{kawai@sci.osaka-cu.ac.jp}}\quad
  Eisaku \textsc{Sakane}\thanks{e-mail:
    \texttt{sakane@sci.osaka-cu.ac.jp}}\quad and\enskip
  Takashi \textsc{Tojo}${}^\ddag$ \\ 
  \noalign{\vskip 2ex} 
  \textit{\small Department of Physics, Osaka City University, 
    Osaka 558-8585, Japan}\\ 
  \textit{\small ${}^\ddag$Up Educational Project, 
    Nishinomiya 663-8204, Japan}}

\date{\today}

\begin{document}

\maketitle

\begin{abstract}
  In Poincar\'e gauge theory of gravity and in
  $\overline{\mbox{Poincar\'e}}$ gauge theory of gravity, we give the
  necessary and sufficient condition in order that the Schwarzschild
  space-time expressed in terms of the Schwarzschild coordinates is
  obtainable as a torsionless exact solution of gravitational field
  equations with a spinless point-like source having the
  energy-momentum density $\widetilde{\mbox{\boldmath
      $T$}}_\mu^{~\nu}(x) =-Mc^2\delta_\mu^{~0}\delta_0^{~\nu}
  \delta^{(3)}(\mbox{\boldmath $x$})$. Further, for the case when this
  condition is satisfied, the energy-momentum and the angular momentum
  of the Schwarzschild space-time are examined in their relations to
  the asymptotic forms of vierbein fields. We show, among other
  things, that asymptotic forms of vierbeins are restricted by
  requiring the equality of the active gravitational mass and the
  inertial mass. Conversely speaking, this equality is violated for a
  class of vierbeins giving the Schwarzschild metric.
\end{abstract}

\section{Introduction}
\label{sec:intro}

There have been many attempts to describe gravity by gauge theory,
among which we have Poincar\'{e}~\cite{HS01} and
$\overline{\mbox{Poincar\'e}}$~\cite{Kawai01} gauge theories. The
former, which shall be written as PGT in short, is formulated by the
requirement of invariance of the theory under the local Poincar\'e
transformations. In this theory, the Poincar\'e group does not work as
a gauge group in the sense of Yang-Mills theories.
$\overline{\mbox{Poincar\'e}}$ gauge theory (\=PGT) is formulated on
the basis of the principal fiber bundle over the space-time possessing
the covering group $\bar{P}_{0}$ of the Poincar\'{e} group as the
structure group, by following the lines of the standard geometric
formulation of Yang-Mills theories as closely as possible, in which
the group $\bar{P}_{0}$ works as the gauge group in the sense of
Yang-Mills theories.

The energy-momentum and angular momentum have been discussed in
Ref.~\citen{HS02} for PGT and in Refs.~\citen{Kawai02} and
\citen{Kawai-Saitoh} for \= PGT, in the general framework of the
formulations, by assuming that the vierbein fields and the Lorentz
gauge fields approach their asymptotic values sufficiently rapidly.

In this paper, we examine these two theories.  We study, among other
things, the following: (1)~We obtain the necessary and sufficient
condition imposed on parameters in gravitational Lagrangian density in
order that the Schwarzschild space-time expressed in terms of the
Schwarzschild coordinates is obtainable as a torsionless exact
solution of the gravitational field equations with a spinless
point-like source located at the origin. (2)~For the case for which
this condition is satisfied, the energy-momentum and the angular
momentum of the Schwarzschild space-time are examined in their
relations to asymptotic behaviors of vierbein fields, and conditions
for vierbeins to give the equality of the active gravitational mass
and the inertial mass are obtained.

\section{Preliminary}
\label{sec:pre}

We briefly summarize the basics of Poincar\'e gauge theory and of 
$\overline{\mbox{Poincar\'e}}$ gauge theory for the convenience of 
latter discussion.

\subsection{Poincar\'e gauge theory}
\label{sec:review-p}

In this theory,~\cite{HS01} the space-time is assumed to be a 
differentiable manifold endowed with the Lorentzian 
metric\footnote{In our conventions, the middle part of the Greek 
alphabet, $\mu, \nu,\lambda,\ldots$, refers to 
0,1,2 and 3, while the initial part, $\alpha, \beta, \gamma,\ldots $,
denotes 1,2 and 3. In a similar way, the middle part of the Latin 
alphabet, $i,j,k,\ldots$, means 0,1,2 and 3, unless otherwise stated. 
While the initial part, $a,b,c,\ldots$, denotes 1,2 and 3.}
$g_{\mu \nu }dx^{\mu}\otimes dx^{\nu }$. 
Here, $\{x^{\mu };\mu =0,1,2,3\}$ is a local 
coordinate of the space-time. Fundamental gravitational field variables 
are vierbeins ${\mbox{\boldmath $e$}}_{k} 
=e^{\mu }_{~k}\partial/\partial x^{\mu }$ and 
Lorentz gauge potentials $A^{kl}_{~~\mu }$. For the duals
${\mbox{\boldmath $e$}}^{k}=e^{k}_{~\mu }dx^{\mu }$ of the vierbeins, 
we have $g_{\mu \nu }=e^{k}_{~\mu }\eta_{kl}e^{l}_{~\nu }$ with 
$(\eta_{kl})\stackrel{\mbox{\scriptsize def}}{=}{\rm diag}(-1,1,1,1)$.
The covariant derivative $D_{k}$ of the field $\varphi $ 
belonging to a representation $\sigma $ of the Lorentz group is given by  
\begin{equation}
\label{eq:cov-p}
D_k \varphi =e^{\mu }_{~k}\left(\partial_{\mu }{\varphi }
+\frac{i}{2}A^{lm }_{~~\mu }M_{lm}\varphi \right)\;,
\end{equation}
where $M_{lm} \stackrel{\mbox{\scriptsize def}}{=}-i\sigma_{*}
(\overline{M}_{lm})$. Here, $\{\overline{M}_{lm}, l,m=0,1,2,3\}$ 
is a basis of the Lie algebra of the Lorentz group satisfying 
the relation
\begin{eqnarray}
\label{eq:alg-Lor} 
[\overline {M}_{kl}, \overline{M}_{mn}]&=&-\eta_{km}\overline{M}_{ln}
-\eta_{ln}\overline{M}_{km}+\eta_{kn}\overline{M}_{lm}+
\eta_{lm}\overline{M}_{kn}\;, \\
\overline{M}_{kl}&=&-\overline{M}_{lk}\;,
\end{eqnarray}
and $\sigma_{*}$ stands for the differential of $\sigma $.

The field strengths of $e^{k}_{~\mu }$ and of $A^{kl}_{~~\mu }$ are
given by
\begin{eqnarray}
  \label{eq:fstr-p}
  T^k_{~lm}&\stackrel{\mbox{\scriptsize def}}{=}&
  e^\mu_{~l}e^\nu_{~m}(\partial_\mu e^k_{~\nu}-\partial_\nu e^k_{~\mu})
  +e^\mu_{~l}A^k_{~m\mu}-e^\mu_{~m}A^k_{~l\mu}\;,\\
  R^{kl}_{~~mn}&\stackrel{\mbox{\scriptsize def}}{=}&
  e^\mu_{~m}e^\nu_{~n}(
  \partial_\mu A^{kl}_{~~\nu}-\partial_\nu A^{kl}_{~~\mu}
  +A^k_{~j\mu}A^{jl}_{~~\nu}-A^k_{~j\nu}A^{jl}_{~~\mu})\;,
\end{eqnarray}
respectively. The gauge potentials $A^{kl}_{~~\mu }$ and the affine 
connection coefficients $\Gamma^{\nu }_{\lambda \mu }$ are related through 
the relation 
\begin{equation}
\label{eq:rel-Lorentz-affine}
A^{k}_{~l\mu }
\equiv \Gamma^{\nu }_{\lambda \mu }e^{k}_{~\nu }e^{\lambda }_{~l}
+e^{k}_{~\nu }\partial_{\mu }e^{\nu }_{l}\;,
\end{equation}
and we have 
\begin{eqnarray}
\label{eq:fsr-tor-cur}
T^{k}_{~\mu \nu }\equiv e^{k}_{~\lambda }T^{\lambda }_{~\mu \nu }\;, 
\; \; \; R^{k}_{~l\mu \nu }
\equiv e^{k}_{~\lambda }e^{\rho }_{~l}R^{\lambda }_{~\rho \mu \nu }
\end{eqnarray}
with
\begin{equation}
\label{eq:tor-cur}
T^{\mu }_{~\nu \lambda }
\stackrel{\mbox{\scriptsize def}}{=}
\Gamma^{\mu }_{\lambda \nu }-
\Gamma^{\mu }_{\nu \lambda }\;, \; \; \; 
R^{\mu }_{~\nu \lambda \rho }\stackrel{\mbox{\scriptsize def}}{=}
\partial_{\lambda }\Gamma^{\mu }_{\nu \rho }
-\partial_{\rho }\Gamma^{\mu }_{\nu \lambda }
+\Gamma^{\mu }_{\tau \lambda }\Gamma^{\tau }_{\nu \rho }
-\Gamma^{\mu }_{\tau \rho }\Gamma^{\tau }_{\nu \lambda }\;.
\end{equation}
The components $T^{\mu }_{~\nu \lambda }$ and 
$R^{\mu }_{~\nu \lambda \rho }$ are those of the torsion tensor and of 
the curvature tensor, respectively, and they are both non-vanishing in 
general. 

The field components $e^{k}_{~\mu }$ and $e^{\mu }_{~k}$ will be used 
to convert Latin and Greek indices. Also, raising and lowering the 
Latin indices are accomplished with the aid of $(\eta^{kl})
\stackrel{\mbox{\scriptsize def}}{=}(\eta_{kl})^{-1}$ and $(\eta_{kl})$,
respectively.

For the matter field $\varphi $, $L_{M}(\varphi ,D_{k}\varphi )$ is a 
Lagrangian\footnote{From now on, Lagrangian density is simply called 
Lagrangian.} invariant under local Lorentz transformations and general
coordinate transformations, if $L_{M}(\varphi ,\partial_{k}\varphi )$ 
is an invariant Lagrangian on the Minkowski space-time.

The gravitational Lagrangian, which is invariant under local Lorentz 
transformations including also inversions and under general
coordinate transformations and at most quadratic in torsion and 
curvature tensors, is given by 
\begin{equation}
\label{eq:grav-lag}
\bar{L}_{G}\stackrel{\mbox{\scriptsize def}}{=}
  L_{T}+L_{R}+aR \\
\end{equation}
with
\begin{eqnarray}
\label{eq:lag-tor-cur}
L_{T}&\stackrel{\mbox{\scriptsize def}}{=}&
  \alpha t^{klm}t_{klm}+\beta v^kv_k+\gamma a^ka_k \;,\\
L_{R}&\stackrel{\mbox{\scriptsize def}}{=}&
  c_1A^{klmn}A_{klmn}+c_2B^{klmn}B_{klmn} \nonumber\\
  & &+c_3C^{klmn}C_{klmn}+c_4E^{kl}E_{kl}+c_5I^{kl}I_{kl}+c_6R^2\;.
\end{eqnarray}
In the above, $\alpha,\beta,\gamma,c_{k}~(k=1,2,\cdots,6)$ and $a$ are real 
constants. Also, $t_{ijk}, v_{k} $ and $a_{k}$ are 
irreducible components of the field strength $T_{klm}$, and  
$A_{klmn}, B_{klmn}, C_{klmn}, E_{kl}$, $I_{kl}$ and $R$ are
irreducible components of the field strength $R_{klmn}$.
These components are given explicitly in Appendix~\ref{appendix:1}.
Then, 
\begin{equation}
\label{eq:action}
\mbox{\boldmath $I$}
\stackrel{\mbox{\scriptsize def}}{=}\frac{1}{c}\int 
\bar{\mbox{\boldmath $L$}}d^{4}x
\end{equation}
is the total action of the system, where $c$ is the light velocity in 
the vacuum and $\bar{\mbox{\boldmath $L$}}$ is defined by
\begin{equation}
\label{eq:total-lag}
\bar{\mbox{\boldmath $L$}}
\stackrel{\mbox{\scriptsize def}}{=}
\sqrt{-g}[\bar{L}_{G}+L_{M}(\varphi, D_{k}\varphi)]
\end{equation}
with $g\stackrel{\mbox{\scriptsize def}}{=}\det(g_{\mu \nu })$.
The parameter $a$ is 
fixed~\cite{HS03}\footnote{The theory has the Newtonian 
limit, also when~\cite{HS03} 
$c_k =\infty~(k=1,2,\cdots, 6)$ and 
$\alpha +4\beta +9\alpha \beta \kappa \\
=0$. We do not deal with this extreme case in the present paper.} 
to be $a=1/2\kappa=c^4/16\pi G$, which we shall assume to hold in 
this paper, by the requirement that the theory has the Newtonian 
limit. Here, $\kappa $ and $G$ stand for the Einstein gravitational 
constant and the Newton gravitational constant, respectively. 

When the torsion is vanishing, $T_{klm}\equiv 0$, the field equations 
$\delta \bar{\mbox{\boldmath $L$}}/\delta e^{i}_{~\mu }=0$ and 
$\delta \bar{\mbox{\boldmath $L$}}/\delta A^{ij}_{~\mu }$ $=0$ 
reduce to\footnote{$A_{\ldots [k\ldots l]\ldots }
\stackrel{\mbox{\scriptsize def}}{=}\frac{1}{2}
(A_{\ldots k\ldots l\ldots }-A_{\ldots l\ldots k\ldots })$.}  
\begin{eqnarray}
\label{eq:fieldeq1}
2aG_{ij}(\{\})+(3c_2+2c_5)\Bigl[
R_{im}(\{\})R_j^{~m}(\{\})+R^{mn}(\{\})
\bigl(R_{imjn}(\{\})&&\nonumber\\
-\frac{1}{2}\eta_{ij}R_{mn}(\{\})\bigr)\Bigr]
-(2c_2+c_5-4c_6)R(\{\})\bigl(R_{ij}(\{\})
-\frac{1}{4}\eta_{ij}R(\{\})\bigr)&=&T_{ij}\;, \\
\label{eq:fieldeq2}
(3c_2+2c_5)\nabla_{[i}G_{j]k}(\{\})
+(c_2+c_5+4c_6)\eta_{k[i}\partial_{j]}G(\{\})=-S_{ijk}\;,
\end{eqnarray}
respectively. Here, $T_{ij}$ and $S_{ijk}$ are the energy-momentum and 
spin densities of the source field $\varphi $, respectively. They are 
defined by 
\begin{eqnarray}
\label{eq:T_ij}
T_{ij}&\stackrel{\mbox{\scriptsize def}}{=}&
\frac{1}{\sqrt{-g}}e_{j\mu}
\frac{\delta\mbox{\boldmath $L$}_M}{\delta e^i_{~\mu}}\;, \\
\label{eq:S_ijk}
S_{ijk} &\stackrel{\mbox{\scriptsize def}}{=}&
-\frac{1}{\sqrt{-g}}e_{k\mu}\frac{\delta\mbox{\boldmath $L$}_M}
{\delta A^{ij}_{~~\mu}} 
\end{eqnarray}
with 
$\mbox{\boldmath $L$}_{M}\stackrel{\mbox{\scriptsize def}}{=}
\sqrt{-g}L_{M}$. Also, $G_{ij}(\{\})$ and $G(\{\})$ stand for the 
Einstein tensor and its trace, respectively:
\begin{equation}
\label{eq:Eistein-tensor-def}
G_{ij}(\{\}) \stackrel{\mbox{\scriptsize def}}{=}
R_{ij}(\{\})-\frac{1}{2}\eta_{ij}R(\{\})\;,\; \; 
G(\{\}) \stackrel{\mbox{\scriptsize def}}{=} \eta^{ij}G_{ij}(\{\})\;,
\end{equation}
where we have defined  
\begin{equation}
\label{eq:R-RC}
R_{ij}(\{\})\stackrel{\mbox{\scriptsize def}}{=}
e^{\mu }_{~i}e^{\nu }_{~j}R^{\lambda }_{~\mu \lambda \nu }(\{\})\;, 
\; \; R(\{\})\stackrel{\mbox{\scriptsize def}}{=}
\eta^{ij}R_{ij}(\{\})
\end{equation}
with the Riemann-Christoffel curvature tensor
\begin{equation}
\label{eq:RC-curv}
R^\lambda_{~\rho\mu\nu}(\{\}) \stackrel{\mbox{\scriptsize def}}{=}
\partial_\mu{\lambda\brace\rho~\nu}
-\partial_\nu{\lambda\brace\rho~\mu}
+{\lambda\brace\sigma~\mu}{\sigma\brace\rho~\nu}
-{\lambda\brace\sigma~\nu}{\sigma\brace\rho~\mu}\;.
\end{equation}

\subsection{$\overline{\mbox{Poincar\'e}}$ gauge theory} 
\label{sec:review-pbar}
This theory~\cite{Kawai01} is formulated on the basis of the principal 
fiber bundle over the space-time possessing the covering group 
$\bar{P}_{0}$ of the proper orthochronous Poincar\'{e} group as the 
structure group. The fundamental field variables are the translational 
gauge potentials $A^{k}_{~\mu }$, the Lorentz gauge potentials 
$A^{kl}_{~~\mu }$, Higgs-type field $\psi =\{\psi^{k}\}$ and matter 
field $\varphi $. This theory is different from Poincar\'{e} gauge 
theory in various respects, among which the following is remarkable:
\begin{description}
\item[(A)] There is the non-dynamical Higgs-type field $\psi
  =\{\psi^{k}\}$ playing key roles in the formulation. Its existence
  is a necessary consequence of a basic postulate on the space-time
  manifold and of the structure of the group $\bar{P}_{0}$.  Also,
  $\psi $ is directly related to the existence of a subbundle which
  gives a spinor structure on the space-time, and the spinor structure
  is built in as a sub-structure of the formulation.  Its field
  equation is automatically satisfied, if those of $A^{k}_{~\mu }$ and
  of $\varphi $ are both satisfied.
\item[(B)] Dual components $e^{k}_{~\mu }$ of vierbeins are related to
  the field $\psi $ and the gauge potentials $A^{k}_{~\mu },
  A^{kl}_{~~\mu }$ through the relation
  \begin{equation}
    \label{eq:relation-pbar}
    e^k_{~\mu}=\partial_\mu\psi^k+A^k_{~l\mu}\psi^{l}+A^k_{~\mu}\;.
  \end{equation}
  The components $e^k_{~\mu}$ are invariant under {\em internal\/}
  translations.
\item[(C)] The field strength $R^{k}_{~\mu \nu }$ of $A^{k}_{~\mu }$
  is given by
  \begin{equation}
    \label{eq:fstr-pbar}
    R^k_{~\mu \nu }\stackrel{\mbox{\scriptsize def}}{=}
    \partial_\mu A^k_{~\nu}-\partial_\nu A^k_{~\mu}
    +A^k_{~l \mu}A^{l}_{\nu }-A^k_{~l \nu }A^{l}_{\mu }\;,
  \end{equation}
  and we have the relation
  \begin{equation}
    \label{eq:fstr-relation}
    T^k_{~\mu \nu }=R^{k}_{~\mu \nu }+R^{k}_{~l \mu \nu }\psi^{l}\;.
  \end{equation}
  The field strengths $T^{k}_{~\mu \nu }$ and $R^{kl}_{~~\mu \nu }$
  are both invariant under {\em internal\/} translations.
\item[(D)] A nonintegrable phase factor describes both of the motions
  of a point and of a vector.~\cite{Kawai03}
\item[(E)] Generators of internal Poincar\'{e} transformations and of
  affine coordinate transformations depend on the choice of the set of
  independent fields variables, i.e., generators for the case when
  $\{\psi^k,A^k_{~\mu},A^{kl}_{~~\mu},\varphi \}$ is chosen as the set
  of independent field variables are different from the corresponding
  ones for the case when $\{\psi^k,e^k_{~\mu},A^{kl}_{~~\mu},\varphi
  \}$ is chosen instead.  When $\{\psi^{k}, A^{k}_{~\mu },
  A^{kl}_{~~\mu }, \varphi \}$ is employed as the set, we have the
  following: For suitable asymptotic forms of field variables at
  spatial infinity, \emph{the conserved total energy-momentum and the
    total\/} (={\em spin}+{\em orbital\/}) \emph{angular momentum of
    an isolated system are obtained as the generators of the internal
    Poincar\'{e} transformations, and the generators of the general
    affine coordinate transformations
    vanish}.~\cite{Kawai02,Kawai-Saitoh}
\item[(F)] It admits the possible existence of matter fields having
  non-vanishing \lq \lq intrinsic" energy-momentum $P_{k}$ (=the
  quantum number associated with the internal
  translation),~\cite{Kawai04} and the covariant derivative of the
  matter field $\varphi $ takes the form
  \begin{equation}
    \label{eq:cov-pbar}
    D_k \varphi =e^{\mu }_{~k}\left(\partial_{\mu }{\varphi }
      +\frac{i}{2}A^{lm }_{~~\mu }M_{lm}\varphi 
      + iA^{l}_{~\mu }P_{l}\varphi \right)\;,
  \end{equation}
  in general.
\item[(G)] An extended new general relativity is obtained as a
  reduction of the theory.~\cite{Kawai-Toma}
\end{description}

The invariance of the action integral under internal $\bar{P}_{0}$
gauge transformations requires the gravitational Lagrangian to be a
function of $T_{klm}$ and of $R_{klmn}$, and the gravitational
Lagrangian agrees with that in PGT\@. Hence, \emph{gravitational field
  equations take the same forms in these theories}.

\section{Schwarzschild space-time}
\label{sec:Schwarzschild}

By Schwarzschild space-time, we mean a space-time having the vanishing
torsion and the Schwarzschild metric which has the expression
\begin{equation}
\label{eq:Schwarzschild-metric}
ds^2=-\left(1-\frac{r_0}{r}\right)(dx^0)^2
+\left(1-\frac{r_0}{r}\right)^{-1}(dr)^2
+r^2\left[(d\theta)^2+\sin^2\theta(d\phi)^2\right]\;,
\end{equation}
in the Schwarzschild coordinates $(x^0,r,\theta,\phi)$. Here, $r_0$ is
the Schwarzschild radius $r_0=2GM/c^2$ with $M$ being the active
gravitational mass of the central gravitating body. A static
chargeless mass point with the active gravitational mass $M$ is
considered to produce the Schwarzschild space-time, and the
energy-momentum density $\widetilde{\mbox{\boldmath $T$}}_\mu^{~\nu}$
of the gravitational source is expected to have the expression
\begin{equation}
\label{eq:emd}
\widetilde{\mbox{\boldmath $T$}}_\mu^{~\nu}(x)
=-Mc^2\delta_\mu^{~0}\delta_0^{~\nu}
\delta^{(3)}(\mbox{\boldmath $x$})\;.
\end{equation}

Both in PGT and in \=PGT, the Schwarzschild space-time is obtainable 
as an exact solution of the gravitational field equations with 
$S_{ijk}\equiv 0$ and $\widetilde{\mbox{\boldmath $T$}}_\mu^{~\nu}(x)
\stackrel{\mbox{\scriptsize def}}{=}
\sqrt{-g}e^k_{~\mu}e^{\nu l}T_{kl}=-Mc^2\delta_\mu^{~0}\delta_0^{~\nu}
\delta^{(3)}(\mbox{\boldmath $x$})$, if~\footnote{In 
Ref.~\citen{HS01}, the following has been shown: (1) For a gravitational 
source with $S_{ijk}\equiv 0$, {\em any\/} solution of the Einstein 
equation is also a solution of the gravitational field equations in 
PGT, if the condition (\ref{eq:parameter-condition}) is satisfied. (2) 
The gravitational field equation 
$\delta \bar{\mbox{\boldmath $L$}}/\delta A^{ij}_{~\mu }=0$ contains 
third derivatives of the metric tensor, unless the condition 
(\ref{eq:parameter-condition}) is satisfied.
\label{page:footnote}} and only if 
\begin{equation}
  \label{eq:parameter-condition}
  3c_2+2c_5=0=c_5+12c_6\;,
\end{equation}
as we shall show below. Since $T_{klm}\equiv 0$ for this space-time, 
the curvature $R_{ijmn}$ agrees with the Riemann-Christoffel curvature 
$R_{ijmn}(\{\})$, and it is given by
\begin{eqnarray}
\label{eq:curvature}
R_{(0)(a)(0)(b)}&=&R_{(0)(a)(0)(b)}(\{\})
=-\frac{1}{2}\frac{h'}{r}\delta_{(a)(b)}
+\left(-\frac{h''}{2}+\frac{1}{2}\frac{h'}{r}\right)
\frac{x^{(a)}x^{(b)}}{r^2}\;,\nonumber\\
R_{(0)(a)(b)(c)}&=&R_{(0)(a)(b)(c)}(\{\})=0 \;,\nonumber\\
R_{(a)(b)(c)(d)}&=&R_{(a)(b)(c)(d)}(\{\})=
\left(\delta_{(a)(c)}\delta_{(b)(d)}
-\delta_{(a)(d)}\delta_{(b)(c)}\right)\frac{h}{r^2} \nonumber \\
& &+\left(\frac{x^{(a)}x^{(c)}}{r^2}\delta_{(b)(d)}
-\frac{x^{(a)}x^{(d)}}{r^2}\delta_{(b)(c)}\right.\nonumber\\
& &\left.~~~~+\frac{x^{(b)}x^{(d)}}{r^2}\delta_{(a)(c)}
-\frac{x^{(b)}x^{(c)}}{r^2}\delta_{(a)(d)}\right)
\left(\frac{1}{2}\frac{h'}{r}-\frac{h}{r^2}\right)
\end{eqnarray}
with $h\stackrel{\mbox{\scriptsize def}}{=}r_0/r$, where we have 
written Lorentz (Latin) indices in parentheses. Substituting 
Eq.~(\ref{eq:curvature}) into the field equation 
(\ref{eq:fieldeq1}), we obtain:
\begin{eqnarray}
\label{eq:emd3}
\widetilde{\mbox{\boldmath $T$}}_0^{~0}
&=&-2a\left(\frac{h'}{r}+\frac{h}{r^2}\right)
+(3c_2+2c_5)\left[\left(\frac{h''}{2}
+\frac{h'}{r}\right)\frac{h''}{2}
-\left(\frac{h'}{r}+\frac{h}{r^2}\right)\frac{h}{r^2}\right]
\nonumber\\
& &+(2c_2+c_5-4c_6)\left[-\left(\frac{h''}{2}
+\frac{h'}{r}\right)^2+\left(\frac{h'}{r}
+\frac{h}{r^2}\right)^2\right] \;,\nonumber\\
\widetilde{\mbox{\boldmath $T$}}_0^{~\alpha}
&=&0=\widetilde{\mbox{\boldmath $T$}}_\alpha^{~0}\;,\nonumber\\
\widetilde{\mbox{\boldmath $T$}}_\alpha^{~\beta}
&=& -2a\left[\delta_\alpha^{~\beta}
\left(\frac{h''}{2}+\frac{h'}{r}\right)-\frac{x^\alpha x^\beta}{r^2}
\left(\frac{h''}{2}-\frac{h}{r^2}\right)\right] \nonumber\\
&& -(3c_2+2c_5)\left(\delta_\alpha^{~\beta}
-2\frac{x^\alpha x^\beta}{r^2}\right)\left[
\left(\frac{h''}{2}+\frac{h'}{r}\right)\frac{h''}{2}
-\left(\frac{h'}{r}+\frac{h}{r^2}\right)\frac{h}{r^{2}}\right]
\nonumber \\
&& +(2c_2+c_5-4c_6)\left(\delta_\alpha^{~\beta}
-2\frac{x^\alpha x^\beta}{r^2}\right)
\left[\left(\frac{h''}{2}+\frac{h'}{r}\right)^2
-\left(\frac{h'}{r}+\frac{h}{r^2}\right)^2\right]\;. 
\end{eqnarray}
We regularize~\cite{Kawai-Sakane} the function $h=r_0/r$ as
$r_0/\sqrt{r^2+\epsilon^2}$, in order to make a distribution
theoretical treatment, then the regularized energy density
$\widetilde{\mbox{\boldmath $T$}}_{0}^{~0}(x;\epsilon )$ of the
gravitational source takes the form
\begin{equation}
\label{eq:emd3-0.0}
\widetilde{\mbox{\boldmath $T$}}_{0}^{~0}(x;\epsilon)
= -\frac{2ar_0\epsilon^2}{r^2(r^2+\epsilon^2)^{3/2}}
+ \Lambda(r;\epsilon)\;.
\end{equation}
Here, we have defined  
\begin{eqnarray}
\label{eq:divergence-term}
\Lambda(r;\epsilon) &\stackrel{\mbox{\scriptsize def}}{=}&
(c_2+c_5+4c_6)r_0^2\left[
\frac{9}{4}\frac{\epsilon^4}{(r^2+\epsilon^2)^5}
-\frac{\epsilon^4}{r^4(r^2+\epsilon^2)^3}\right]\nonumber\\
&& -(3c_2+2c_5)r_0^2\left[
\frac{3}{2}\frac{\epsilon^2}{(r^2+\epsilon^2)^4}
+\frac{\epsilon^2}{r^2(r^2+\epsilon^2)^3}\right]\;.
\end{eqnarray}
The first term in the right hand side (r.h.s.) of 
Eq.~(\ref{eq:emd3-0.0}) has the well-defined limit,
\begin{equation}
\label{eq:limit}
\lim_{\varepsilon \rightarrow 0}
\left\{-\frac{2ar_0\epsilon^2}{r^2(r^2+\epsilon^2)^{3/2}}\right\}
=-Mc^{2}\delta^{(3)}(\mbox{\boldmath $x$})\;,
\end{equation}
where we have used $a=c^{4}/16\pi G$. For the second term 
$\Lambda (r;\varepsilon)$, we have 
\begin{equation}
\label{eq:dist-int}
\int_0^\infty\!\Lambda(r;\epsilon)r^2dr
= -\frac{r_0^2}{\epsilon^3}(c_2+c_5+4c_6)
\int_0^\infty\frac{dx}{x^2}
+\frac{15\pi}{1024}\frac{r_0^2}{\epsilon^3}(19c_2+35c_5+268c_6)\;,
\end{equation}
in which the integral in the r.h.s.~is diverging. As is known from 
Eqs.~(\ref{eq:emd3}), (\ref{eq:emd3-0.0}), (\ref{eq:limit}) and 
(\ref{eq:dist-int}), we have the limit 
\begin{equation}
\label{eq:limit-energy-density}
\widetilde{\mbox{\boldmath $T$}}_{\mu }^{~\nu }(x)=
\lim_{\epsilon \rightarrow 0}
\widetilde{\mbox{\boldmath $T$}}_{\mu }^{~\nu }(x;\epsilon)
=-Mc^{2}\delta_{\mu }^{~0}\delta_{0}^{~\nu }
\delta^{(3)}(\mbox{\boldmath $x$})\;,
\end{equation}
if and only if the condition  
\begin{equation}
  \label{eq:parameter-condition2}
  c_2+c_5+4c_6=0=19c_2+35c_5+268c_6\;,
\end{equation}
which is equivalent to Eq.~(\ref{eq:parameter-condition}), is
satisfied.\footnote{As for energy-momentum densities of the
  gravitational field, see \S 5.~for PGT and \S 6.~for \= PGT,
  respectively.}  Obviously, the field equation (\ref{eq:fieldeq2})
with $S_{ijk}\equiv 0$ is satisfied identically, if the condition
(\ref{eq:parameter-condition}) is satisfied.\cite{HS01}

From the above discussion and from the fact mentioned in the footnote
on page \pageref{page:footnote}, we know that the Lagrangian $L_{R}$
with the condition (\ref{eq:parameter-condition}), which we shall
employ in the following, is favorable in various respects.

\section{Spherically symmetric vierbeins giving the \\
  Schwarzschild metric}
\label{sec:vierbeins}

In view of the fact that the Schwarzschild space-time is
asymptotically Minkowskian,~\cite{HS02} we choose vierbeins satisfying
\begin{equation}
\label{eq:asympt-vier}
\lim_{r\rightarrow \infty}e^k_{~\mu}=e^{(0)k}_{~~~~\mu}
\end{equation}
with $e^{(0)k}_{~~~~\mu}$ being constants satisfying 
$e^{(0)k}_{~~~~\mu}\eta_{kl}e^{(0)l}_{~~~~\nu}=\eta_{\mu \nu }$.
The general forms of components $e^{k}_{~\mu }$ having spherical 
symmetry have been given by Robertson,~\cite{Robertson} which can be 
written as
\begin{eqnarray}
\label{eq:vierbein}
e^{(0)}_{~~~0}&=&A \;,\quad
e^{(0)}_{~~~\alpha}=B\frac{x^\alpha}{r} \;,\quad
e^{(a)}_{~~~0}=C\frac{x^{(a)}}{r} \;,\nonumber \\
e^{(a)}_{~~~\alpha}&=&D\delta^{(a)}_{~~~\alpha}
+E\frac{x^{(a)}x^{\alpha}}{r^2}
+F\epsilon_{(a)\alpha\beta}\frac{x^\beta}{r}\;.
\end{eqnarray}
Here, $A,B,C,D,E$ and $F$ are functions of 
$r=\sqrt{(x^1)^2+(x^2)^2+(x^3)^2}$ 
and of $x^0$, and $\epsilon_{(a)(b)(c)}$ stands for the 
three-dimensional Levi-Civita symbol with $\epsilon_{(1)(2)(3)}=1$.
These give the Schwarzschild metric (\ref{eq:Schwarzschild-metric}), 
if and only if 
\begin{eqnarray}
\label{eq:AF-condition}
A^2-C^2 &=& 1-h\;,\quad AB-C(D+E)=0\;,\quad D^2+F^2=1\;,\nonumber\\
-B^2+(D+E)^2 &=& (1-h)^{-1}\;.
\end{eqnarray}
The six functions $A,B,\cdots $ and $F$ are required to satisfy the
four relations in Eq.~(\ref{eq:AF-condition}). Thus, if $A$ and $D$ are 
fixed, the others are uniquely determined up to the sign.
Equations (\ref{eq:asympt-vier}), (\ref{eq:vierbein}) and 
(\ref{eq:AF-condition}) lead to 
\begin{eqnarray}
\label{eq:asympt-AF}
\lim_{r\to\infty}A&=&A_\infty\;,\quad
\lim_{r\to\infty}D=D_\infty\;,\nonumber\\
\lim_{r\to\infty}B &=& \lim_{r\to\infty}C
=\lim_{r\to\infty}E=\lim_{r\to\infty}F=0
\end{eqnarray}
with $A_\infty $ and $D_\infty$ being constants such that
\begin{equation}
\label{eq:AD-condition}
(A_\infty)^2=1=(D_\infty)^2\;.
\end{equation}
From Eq.~(\ref{eq:AF-condition}), it follows that 
\begin{equation}
\label{eq:BDE}
B(1-h)=\Delta C\;, \; \; (D+E)(1-h)=\Delta A\;,\; \; \Delta^{2}=1 
\end{equation}
with $\Delta \stackrel{\mbox{\scriptsize def}}{=}A(D+E)-BC$.

In what follows, we shall choose, for simplicity, the gauge of 
internal Lorentz group in a way such that $A$ and hence $B$ and $C$ 
are independent of~\footnote{This gauge choice is always possible 
for an arbitrarily given spherically symmetric 
vierbeins giving the Schwarzschild metric.} $x^{0}$.

In Appendix~\ref{appendix:2}, we give an example of the set of 
spherically symmetric vierbeins which gives the Schwarzschild metric 
and has a $x^{0}$-independent $A$.

\section{Generators in Poincar\'e gauge theory}
\label{sec:gen-p}

In this section, following Hayashi and Shirafuji,~\cite{HS02} we 
examine generators of Lorentz gauge transformations and of 
Poincar\'{e} coordinate transformations in the Schwarzschild 
space-time. We employ, instead of $\bar{\mbox{\boldmath $L$}}_G$, 
the following $\mbox{\boldmath $L$}_G$ 
\begin{equation}
\label{eq:gravi-lag2-p}
\mbox{\boldmath $L$}_G \stackrel{\mbox{\scriptsize def}}{=}
\bar{\mbox{\boldmath $L$}}_G
+2a\partial_\nu(\sqrt{-g}\;e^{\mu}_{~k}e^{\nu}_{~l}A^{kl}_{~~\mu})\;.
\end{equation}
as the gravitational Lagrangian. Thus, the total Lagrangian becomes 
\begin{equation}
\label{eq:lagrangian2}
\mbox{\boldmath $L$} \stackrel{\mbox{\scriptsize def}}{=}
\mbox{\boldmath $L$}_G+\mbox{\boldmath $L$}_M\;.
\end{equation}
The divergence term in the r.h.s.~of Eq.~(\ref{eq:gravi-lag2-p}) is to 
get reasonable generators,~\cite{HS02,Kawai02} and it does not 
affect field equations. 

The canonical energy-momentum $M_{\mu }$, the orbital angular momentum
$L^{\mu \nu }$ and the spin angular momentum $S_{ij}$ are generators
of coordinate translations, of Lorentz coordinate transformations and
of {\em internal\/} Lorentz transformations, respectively, and they
are expressed as,~\cite{HS02}\footnote{In general, integrands
  in representations (\ref{eq:M-def-p}), (\ref{eq:L-def-p}),
  (\ref{eq:S-def-p}), etc.~of physical quantities are singular at
  $\mbox{\boldmath $x$}=\mbox{\boldmath $0$}$ and on the horizon
  $r=r_{0}$. But, for the case of the example given in
  Appendix~\ref{appendix:2}, this cause no trouble, if $K(x^{0})={\rm
    constant}>0$, For the energy-momentum $M_{\mu }$, for example,
  this can be confirmed easily by using Eqs.~(\ref{eq:M-def-p}),
  (\ref{eq:identity-p}), (\ref{eq:super-Psi2}), (\ref{eq:A}) $\sim $
  (\ref{eq:E}). In what follows, we restrict ourselves to cases such
  that the singularities at $\mbox{\boldmath $x$}=\mbox{\boldmath
    $0$}$ and on the horizon $r=r_{0}$ cause no trouble (see also the
  footnote \ref{page:footnote2} on page \pageref{page:footnote2}).}
\begin{eqnarray}
\label{eq:M-def-p}
M_\mu &\stackrel{\mbox{\scriptsize def}}{=}& 
\int_\sigma{}^{\rm tot}\widetilde{\mbox{\boldmath $T$}}_\mu^{~\nu}
d\sigma_\nu\;,\\
\label{eq:L-def-p}
L^{\mu \nu} &\stackrel{\mbox{\scriptsize def}}{=}&
\eta^{\nu \rho }\int_\sigma\left(
x^\mu~{}^{\rm tot}\widetilde{\mbox{\boldmath $T$}}_{\rho }^{~\lambda }
-{\mbox{\boldmath $\Psi $}}_{\rho }^{~\mu \lambda }\right)
d\sigma_{\lambda }-(\mu \rightleftarrows \nu)\;,\\
\label{eq:S-def-p}
S_{ij} &\stackrel{\mbox{\scriptsize def}}{=}&
\int_\sigma{}^{\rm tot}\widetilde{\mbox{\boldmath $S$}}_{ij}^{~~\nu}
d\sigma_\nu
\end{eqnarray}
with
\begin{eqnarray}
\label{eq:EMD-def-p}
{}^{\rm tot}\widetilde{\mbox{\boldmath $T$}}_\mu^{~\nu}
&\stackrel{\mbox{\scriptsize def}}{=}&
{\widetilde{\mbox{\boldmath $T$}}}^{\prime~\nu}_{\mu}
+{\widetilde{\mbox{\boldmath $t$}}}_{\mu }^{~\nu}\;,\\
\label{eq:super-po1-p}
{\mbox{\boldmath $\Psi $}}_\mu^{~\nu\rho} 
&\stackrel{\mbox{\scriptsize def}}{=}&
\frac{\partial\mbox{\boldmath $L$}}{\partial e^k_{~\nu,\rho}}
e^k_{~\mu}
+\frac{\partial\mbox{\boldmath $L$}}{\partial A^{kl}_{~~\nu,\rho}}
A^{kl}_{~~\mu}\;,\\
\label{eq:SD-def-p}
{}^{\rm tot}\widetilde{\mbox{\boldmath $S$}}_{ij}^{~~\nu}
&\stackrel{\mbox{\scriptsize def}}{=}&
-2\frac{\partial\mbox{\boldmath $L$}}
{\partial e^{[i}_{~~\lambda, \nu }}e_{j]\lambda}
-4\frac{\partial\mbox{\boldmath $L$}}
{\partial A^{[ik}_{~~~\lambda, \nu }}{{A_{j]}}^{k}}_{\lambda}
-i\frac{\partial\mbox{\boldmath $L$}}{\partial\varphi_{,\mu}}
M_{ij}\varphi\;. 
\end{eqnarray}
Here, $\sigma $ and $d\sigma_{\nu }$ stand for a space-like surface 
and the surface element on it, respectively, and we have 
defined\footnote{The gravitational energy-momentum density 
${\widetilde{\mbox{\boldmath $t$}}}_{\mu }^{~\nu}$ is not vanishing, 
even if the condition (\ref{eq:parameter-condition}) is satisfied.  
The integrated gravitational energy-momentum, however, 
vanishes, when the conditions (\ref{eq:parameter-condition}) 
and (\ref{eq:AD-condition-p}) are both satisfied.
This density does not reduce to the gravitational energy-momentum 
density employed in Ref.~\citen{Kawai-Sakane}, even for the case with 
$c_{k}=0\; (k=1,2,\cdots ,6)$ and with $T_{klm}\equiv 0$ for which the 
relation $\bar{L}_{G}=aR(\{\})$ holds.}
\begin{eqnarray}
\label{eq:energy-momentum-density-1}
{\widetilde{\mbox{\boldmath $T$}}}^{\prime~\nu}_{\mu }
&\stackrel{\mbox{\scriptsize def}}{=}&
\delta_\mu^{~\nu}\mbox{\boldmath $L$}_{M}
-\frac{\partial\mbox{\boldmath $L$}_{M}}{\partial\varphi_{,\nu}}
\varphi_{,\mu} \;,\\
\label{eq:energy-momentum-density-2}
{\widetilde{\mbox{\boldmath $t$}}}_{\mu }^{~\nu}
&\stackrel{\mbox{\scriptsize def}}{=}&
\delta_\mu^{~\nu}\mbox{\boldmath $L$}_{G}
-\frac{\partial\mbox{\boldmath $L$}_{G}}{\partial e^k_{~\lambda,\nu}}
e^k_{~\lambda,\mu}
-\frac{\partial\mbox{\boldmath $L$}_{G}}{\partial A^{kl}_{~~\lambda,\nu}}
A^{kl}_{~~\lambda,\mu}\;.
\end{eqnarray}
There is the relation, 
\begin{equation}
\label{eq:equality}
{\widetilde{\mbox{\boldmath $T$}}}_{\mu }^{\prime~\nu}=
{\widetilde{\mbox{\boldmath $T$}}}_{\mu }^{~\nu}\;,
\end{equation}
which follows from the fact that $\mbox{\boldmath $L$}_{M}$ is a 
scalar density.
Also, we have the identities
\begin{eqnarray}
\label{eq:identity-p}
-\frac{\delta\mbox{\boldmath $L$}}{\delta e^k_{~\nu}}e^k_{~\mu}
-\frac{\delta\mbox{\boldmath $L$}}{\delta A^{kl}_{~~\nu}}A^{kl}_{~~\mu}
+{}^{\rm tot}\widetilde{\mbox{\boldmath $T$}}_\mu^{~\nu} &\equiv&
\partial_{\lambda}
{\mbox{\boldmath $\Psi $}}_{\mu }^{~\nu \lambda}\;,\\
\label{eq:identity-p2}
\frac{\delta\mbox{\boldmath $L$}}{\delta A^{ij}_{~~\mu}}
+\frac{1}{2}{}^{\rm tot}\widetilde{\mbox{\boldmath $S$}}_{ij}^{~~\mu}
&\equiv& 
\frac{1}{2}\partial_{\nu }{\mbox{\boldmath $\Sigma $}}_{ij}^{~~\mu \nu}\;,
\end{eqnarray}
where we have defined 
\begin{equation}
\label{eq:super-po2-p}
{\mbox {\boldmath $\Sigma $}}_{ij}^{~~\mu \nu} 
\stackrel{\mbox{\scriptsize def}}{=}
-2\left[\frac{\partial\mbox{\boldmath $L$}}
{\partial A^{ij}_{~~\mu,\nu}}
-2a(\sqrt{-g}e^{[\mu }_{~~i}e^{\nu ]}_{~~j}
-e^{(0)[\mu }_{~~~~~i}e^{(0)\nu ]}_{~~~~~j})
\right]\;.
\end{equation}
More explicitly, the superpotential 
${\mbox{\boldmath $\Psi $}}_{\mu }^{~\nu \rho }$ is expressed as
\begin{eqnarray}
\label{eq:super-Psi}
{\mbox{\boldmath $\Psi $}}_{\mu }^{~\nu \rho }&=&
-(2a/\sqrt{-g})g_{\mu \lambda }\partial_\sigma
(gg^{\lambda [\nu }g^{\rho ]\sigma}) \nonumber \\
& &+2a\sqrt{-g}\bigl[
\delta_\mu^{~[\nu}(e^{\rho]}_{~~i}e^{\lambda i}_{~~,\lambda}
-e^\lambda_{~i}e^{\rho ]i}_{~~~,\lambda})
+e^{[\nu}_{~~i}e^{\rho ]i}_{~~~,\mu}\bigr] \nonumber \\
& &-2\mbox{\boldmath $J$}^{[ij][\nu \rho]}\Delta_{ij\mu}\;,
\end{eqnarray}
where we have defined 
\begin{eqnarray}
\label{eq:J-P}
  \mbox{\boldmath $J$}^{ij\nu \rho}
  &\stackrel{\mbox{\scriptsize def}}{=}&
  -\frac{1}{2}\frac{\partial\mbox{\boldmath $L$}_R}
  {\partial A_{ij\nu,\rho}}\;,\\
\label{eq:rot-coeff}
\Delta_{ij\mu }&\stackrel{\mbox{\scriptsize def}}{=}&
\frac{1}{2}e^{k}_{~\mu }(C_{ijk}-C_{jik}-C_{kij})
\end{eqnarray}
with 
\begin{equation}
\label{eq:C_ijk}
C_{ijk}\stackrel{\mbox{\scriptsize def}}{=}
e^{\nu }_{~j}e^{\lambda }_{~k}
(\partial_{\nu }e_{i\lambda }-\partial_{\lambda }e_{i\nu })\;.
\end{equation}  
For ${\mbox{\boldmath $\Sigma $}}_{ij}^{~~\mu \nu}$, we have 
\begin{equation}
\label{eq:super-Sigma}
{\mbox{\boldmath $\Sigma $}}_{ij}^{~~\mu \nu}
=4a\left[\sqrt{-g}e^{[\mu }_{~~i}e^{\nu ]}_{~~j}
-e^{(0)[\mu }_{~~~~~i}e^{(0)\nu ]}_{~~~~~j}\right]
+4\mbox{\boldmath $J$}_{[ij]}^{~~~[\mu \nu]}\;.
\end{equation}
The expression of ${\mbox{\boldmath $\Psi $}}_{0}^{~0 \alpha}$ and of 
${\mbox{\boldmath $\Psi $}}_{\alpha}^{~0 \beta }$
in terms of $A,B,\cdots $ and $F$ is given by
\begin{eqnarray}
\label{eq:super-Psi2}
{\mbox{\boldmath $\Psi $}}_{0}^{~0\alpha} &=&
4a\frac{1}{r}\left(1-h-\frac{AD}{\Delta}\right)\frac{x^\alpha}{r}
+3c_2\frac{hh'}{r^2}
\frac{x^\alpha}{r} \;,\nonumber\\
{\mbox{\boldmath $\Psi $}}_\alpha^{~0\beta} &=&
2a\left[A'B-C'(D+E)\right]\left(\delta_\alpha^\beta
-\frac{x^\alpha x^\beta}{r^2}\right)
-2a\frac{1}{r}\frac{BD}{\Delta}\left(\delta_\alpha^\beta
+\frac{x^\alpha x^\beta}{r^2}\right) \nonumber\\
&& +2a\frac{1}{r}\frac{BF}{\Delta}
\epsilon_{\alpha\beta\gamma}\frac{x^\gamma}{r}
+3c_2\frac{h'}{r^2}\left[\frac{BD}{\Delta}\left(\delta_\alpha^\beta
-\frac{x^\alpha x^\beta}{r^2}\right)+\frac{BF}{\Delta}
\epsilon_{\alpha\beta\gamma}\frac{x^\gamma}{r}\right] \nonumber\\
& &-6c_2\frac{h}{r^2}\left[A'B-C'(D+E)\right]
\frac{x^\alpha x^\beta}{r^2}\;,
\end{eqnarray}
where we have defined 
$h'\stackrel{\mbox{\scriptsize def}}{=}dh/dr,
A'\stackrel{\mbox{\scriptsize def}}{=}dA/dr$, etc. Also, we have 
\begin{eqnarray}
\label{eq:super-Sigma2}
{\mbox{\boldmath $\Sigma $}}_{(0)(a)}^{~~~~~~0\alpha} &=&
2a\left[\frac{F^2-DE}{\Delta}\left(\delta_{(a)}^\alpha
-\frac{x^\alpha x^{(a)}}{r^2}\right)
-\frac{(D+E)F}{\Delta}\epsilon_{(a)\alpha\beta}
\frac{x^\beta}{r}\right] \nonumber\\
& &-6c_2\biggl[\frac{1}{2}\frac{h'}{r}\frac{D(D+E)}{\Delta}
\left(\delta^\alpha_{(a)}-\frac{x^\alpha x^{(a)}}{r^2}\right) \nonumber\\
& &~~~~\quad+\frac{1}{2}\frac{h'}{r}\frac{(D+E)F}{\Delta}
\epsilon_{(a)\alpha\beta}\frac{x^\beta}{r}
+\frac{1}{\Delta}\frac{h}{r^2}\frac{x^\alpha x^{(a)}}{r^2}\biggr]\;,
\nonumber \\
{\mbox{\boldmath $\Sigma $}}_{(a)(b)}^{~~~~~~0\alpha} &=& 
\left(2a+3c_2\frac{h'}{r}\right)
\left[\frac{BD}{\Delta}\left(
\delta^\alpha_{(a)}\frac{x^{(b)}}{r}
-\delta^\alpha_{(b)}\frac{x^{(a)}}{r}\right) \right. \nonumber \\
& &\left.\qquad\qquad\qquad\quad+\frac{BF}{\Delta}
\left(\epsilon_{(a)\alpha\beta}\frac{x^\beta x^{(b)}}{r^2}
-\epsilon_{(b)\alpha\beta}\frac{x^\beta x^{(a)}}{r^2}\right)\right]\;.
\end{eqnarray}

\subsection{Energy-Momentum}
\label{sec:em-p}

Equation (\ref{eq:M-def-p}) can be written as
\begin{equation}
\label{eq:em-psi}
M_{\mu }
=\int{\mbox{\boldmath $\Psi $}}_{\mu}^{~0\alpha}r^2n_\alpha d\Omega \;,
\end{equation}
with the aid of Eq.~(\ref{eq:identity-p}), where $d\Omega $ stands 
for the differential solid angle. We can show that
\begin{eqnarray}
\label{eq:em2-p}
M_0&=&16\pi a\lim_{r\to\infty}
r\left(1-h-\frac{AD}{\Delta}\right)\;,\nonumber \\
M_\alpha&=&0\;,
\end{eqnarray}
by the use of Eq.~(\ref{eq:super-Psi2}). We look for the 
condition imposed on the functions $A, B, C, D, E$ and $F$ 
by the requirement 
\begin{equation}
\label{eq:momentum}
M_\mu=-\delta_\mu^{~0}Mc^2\;,
\end{equation}
which implies the equality of the active gravitational mass and the
inertial mass.  The condition (\ref{eq:momentum}) is equivalent to
\begin{equation}
\label{eq:AD-condition-p}
\lim_{r\to\infty}r\left(1-\frac{AD}{\Delta}\right)=\frac{r_0}{2}\;.
\end{equation}
If we express $A/A_{\infty }$ and $D/D_{\infty }$ as 
\begin{equation}
\label{eq:case-p}
\frac{A}{A_{\infty }}
=1-\frac{h}{2}+P\;,\; \; \frac{D}{D_{\infty }}=1+Q\;, 
\end{equation} 
then the relation $\lim_{r\to \infty }P=0=\lim_{r\to \infty }Q$
follows.  Equation (\ref{eq:AD-condition-p}) is equivalent to the
condition
\begin{equation}
\label{eq:YP-condition}
\lim_{r\to \infty }r(P+Q+PQ)=0\;.
\end{equation}
Equation~(\ref{eq:momentum}) agrees with Eq.~(5$\cdot $6) of
Ref.~\citen{HS02}, which has been obtained for generic systems being
at rest as a whole.

\subsection{Angular Momentum}
\label{sec:angular-p}
By virtue of Eqs.~(\ref{eq:identity-p}) and (\ref{eq:identity-p2}),
Eqs.~(\ref{eq:L-def-p}) and (\ref{eq:S-def-p}) can be rewritten as 
\begin{eqnarray}
\label{eq:angular-p}
L^{\mu \nu}&=&
\eta^{\nu \lambda}\int x^\mu 
{\mbox{\boldmath $\Psi $}}_{\lambda }^{~0\alpha}r^2n_\alpha d\Omega
- (\mu \rightleftarrows \nu)\;,\\
\label{eq:spin-angular-p}
S_{ij} &=&
\int{\mbox {\boldmath $\Sigma $}}_{ij}^{~~0\alpha}r^2 n_\alpha 
d\Omega \;.
\end{eqnarray}
These lead to, upon using Eqs.~(\ref{eq:super-Psi2}) and 
(\ref{eq:super-Sigma2}), 
\begin{equation}
\label{eq:angular2-p}
L^{\mu \nu}=0\;,\quad S_{ij}=0\;,
\end{equation}
which hold for any $A,B,C,D,E$ and $F$ satisfying 
the conditions (\ref{eq:AF-condition}) and (\ref{eq:AD-condition}).

\section{Generators in $\overline{\mbox{Poincar\'e}}$ gauge theory}
\label{sec:gen-pbar}
In this section, on the basis of the discussion in
Refs.~\citen{Kawai02} and \citen{Kawai-Saitoh}, we examine generators
of internal Poincar\'{e} transformations and of general affine
coordinate transformations for the Schwarzschild space-time in \=PGT.

\subsection{The case when 
$\{\psi^k,A^k_{~\mu},A^{kl}_{~~\mu},\varphi \}$ 
is employed as the set of independent field variables}
\label{sec:choice1-pbar}

Let us denote the Lagrangians $\mbox{\boldmath $L$}$ and
$\mbox{\boldmath $L$}_{G}$ expressed as functions of
$\psi^k,A^k_{~\mu},A^{kl}_{~~\mu}$, $\varphi $ and of their
derivatives by $\hat{\mbox{\boldmath $L$}}$ and $\hat{\mbox{\boldmath
    $L$}}_{G}$, respectively. For this case, the generator
$\hat{M}_{k}$ of {\em internal\/} translations and the generator
$\hat{S}_{kl}$ of {\em internal\/} Lorentz transformations
are~\cite{Kawai02} the dynamical energy-momentum and the total
($=$spin$+$orbital) angular momentum, respectively, and they are
expressed as
\begin{eqnarray}
\label{eq:M-def-pbar}
\hat{M}_k &\stackrel{\mbox{\scriptsize def}}{=}&
\int_\sigma{}^{\rm tot}\hat{\mbox{\boldmath $T$}}_k^{~\mu}
d\sigma_\mu\;,\\
\label{eq:S-def-pbar}
\hat{S}_{kl} &\stackrel{\mbox{\scriptsize def}}{=}&
\int_\sigma{}^{\rm tot}\hat{\mbox{\boldmath $S$}}_{kl}^{~~\mu}
d\sigma_\mu 
\end{eqnarray}
with
\begin{eqnarray}
\label{eq:MS-def2-pbar}
{}^{\rm tot}\hat{\mbox{\boldmath $T$}}_k^{~\mu}
&\stackrel{\mbox{\scriptsize def}}{=}&
\frac{\partial\hat{\mbox{\boldmath $L$}}}{\partial\psi^k_{~,\mu}}
+i\frac{\partial\hat{\mbox{\boldmath $L$}}}
{\partial\varphi_{,\mu}}P_k\varphi
+\frac{\partial\hat{\mbox{\boldmath $L$}}}
{\partial A^l_{~\nu,\mu}}A_{k~\nu}^{~l}\;,\\
{}^{\rm tot}\hat{\mbox{\boldmath $S$}}_{kl}^{~~\mu}
&\stackrel{\mbox{\scriptsize def}}{=}&
-2\left(
\frac{\partial\hat{\mbox{\boldmath $L$}}}
{\partial\psi^{[k}_{~~,\mu}}\psi_{l]}
+\hat{\mbox{\boldmath $F$}}_{[k}^{~~\nu\mu}A_{l]\nu}
+2\hat{\mbox{\boldmath $F$}}_{[km}^{~~~~\nu\mu}A_{l]~\nu}^{~m}+
\frac{i}{2}
\frac{\partial\hat{\mbox{\boldmath $L$}}}{\partial\varphi_{,\mu}}
M_{kl}\varphi\right)\;,\\
\hat{\mbox{\boldmath $F$}}_k^{~\mu\nu} 
&\stackrel{\mbox{\scriptsize def}}{=}&
\frac{\partial\hat{\mbox{\boldmath $L$}}}{\partial A^k_{~\mu,\nu}}\;,
\quad \hat{\mbox{\boldmath $F$}}_{kl}^{~~\mu\nu}
\stackrel{\mbox{\scriptsize def}}{=}
\frac{\partial\hat{\mbox{\boldmath $L$}}}
{\partial A^{kl}_{~~\mu,\nu}}\;,\\
\hat{\mbox{\boldmath $\Sigma$}}_{kl}^{~~\mu\nu} 
&\stackrel{\mbox{\scriptsize def}}{=}&
-2\left[\hat{\mbox{\boldmath $F$}}_{kl}^{~~\mu\nu}
-2a\left(\sqrt{-g}e^{[\mu}_{~~k}e^{\nu]}_{~~l}
-e^{(0)[\mu}_{~~~~~k}e^{(0)\nu]}_{~~~~~l}\right)\right]\;.
\end{eqnarray}
Also, there are the identities 
\begin{eqnarray}
\label{eq:identity-pbar}
-\frac{\delta\hat{\mbox{\boldmath $L$}}}{\delta A^k_{~\mu}}
+{}^{\rm tot}\hat{\mbox{\boldmath $T$}}_k^{~\mu} &\equiv& 
\partial_\nu\hat{\mbox{\boldmath $F$}}_k^{~\mu\nu}\;,\\
\label{eq:identity-pbar2}
\frac{\delta\hat{\mbox{\boldmath $L$}}}{\delta A^{kl}_{~~\mu}}
+\frac{1}{2}{}^{\rm tot}\hat{\mbox{\boldmath $S$}}_{kl}^{~~\mu}
&\equiv& 
\frac{1}{2}\partial_\nu
\hat{\mbox{\boldmath $\Sigma$}}_{kl}^{~~\mu\nu}\;.
\end{eqnarray}
The functions $\hat{\mbox{\boldmath $F$}}_k^{~\mu \nu}$ and 
$\hat{\mbox{\boldmath $F$}}_{kl}^{~~\mu \nu}$ have the expressions 
\begin{eqnarray} 
\label{eq:super-F-pbar} 
\hat{\mbox{\boldmath $F$}}_k^{~\mu\nu}
&=& (2a/\sqrt{-g})e_{k\rho}\partial_\sigma
\left[(-g)g^{[\mu\rho}g^{\nu]\sigma}\right] \nonumber\\
&& +2a\sqrt{-g}\left[e^{[\mu}_{~~k}
\left(e^{\nu]l}e^\lambda_{~l,\lambda}
-e^{\lambda l}e^{\nu]}_{~~l,\lambda}\right)
+e^\lambda_{~k}e^{[\mu l}e^{\nu] }_{~~l,\lambda}\right]\;,
\nonumber \\
\label{eq:super-F-pbar2}
\hat{\mbox{\boldmath $F$}}_{kl}^{~~\mu\nu}
&=& -2\hat{\mbox{\boldmath $J$}}_{[kl]}^{~~~[\mu\nu]}
+\hat{\mbox{\boldmath $F$}}_{[k}^{~~\mu\nu}\psi_{l]}
\end{eqnarray}
with 
\begin{equation}
\label{eq:J-pbar}
\hat{\mbox{\boldmath $J$}}_{[kl]}^{~~~[\mu\nu]}
\stackrel{\mbox{\scriptsize def}}{=}
-\frac{1}{2}\frac{\partial\hat{\mbox{\boldmath $L$}}_R}
{\partial A^{kl}_{~~\mu,\nu}}\;.
\end{equation}
The expressions of $\hat{\mbox{\boldmath $F$}}_{(0)}^{~~~0\alpha}$ 
and of $\hat{\mbox{\boldmath $F$}}_{(a)}^{~~~0\alpha} 
$ in terms of $A,B,\cdots $ and $F$ are given by 
\begin{eqnarray}
\label{eq:super-F2-pbar}
\hat{\mbox{\boldmath $F$}}_{(0)}^{~~~0\alpha}
&=& 4a\frac{1}{r}\left(A-\frac{D}{\Delta }\right)\frac{x^\alpha}{r}
\nonumber\;, \\
\hat{\mbox{\boldmath $F$}}_{(a)}^{~~~0\alpha}
&=& -4a\frac{1}{r}C\frac{x^{(a)}x^\alpha}{r^2}+
2a\left\{D\left[A'B-C'(D+E)\right]-\frac{1}{r}\frac{B}{\Delta}
\right\}\left(\delta^\alpha_{(a)}
-\frac{x^{(a)}x^\alpha}{r^2}\right)\nonumber\\
&& +2aF\left[A'B-C'(D+E)\right]
\epsilon_{(a)\alpha\beta}\frac{x^\beta}{r}\;.
\end{eqnarray}
For $\hat{\mbox{\boldmath $\Sigma$}}_{kl}^{~~\mu\nu}$, we have  
\begin{eqnarray}
\label{eq:super-Sigma2-pbar}
\hat{\mbox{\boldmath $\Sigma$}}_{(0)(a)}^{~~~~~~0\alpha}
&=& 2a\left[\frac{1}{\Delta}(F^{2}-DE)\left(\delta^\alpha_{(a)}
-\frac{x^\alpha x^{(a)}}{r^2}\right)-\frac{(D+E)F}{\Delta}
\epsilon_{(a)\alpha\beta}\frac{x^\beta}{r}\right]\nonumber\\
&& -3c_2\Biggl[\frac{h'}{r}\frac{D(D+E)}{\Delta}
\left(\delta^\alpha_{(a)}-\frac{x^\alpha x^{(a)}}{r^2}\right)
+\frac{h'}{r}\frac{(D+E)F}{\Delta}
\epsilon_{(a)\alpha\beta}\frac{x^\beta}{r} \nonumber\\
&& +\frac{2}{\Delta}\frac{h}{r^2}\frac{x^\alpha x^{(a)}}{r^2}\Biggr]
-4a\frac{1}{r}\left(A-\frac{D}{\Delta}\right)
\frac{x^\alpha}{r}\psi_{(a)} \nonumber\\
&& +2a\Biggl(-\frac{2}{r}C\frac{x^\alpha x^{(a)}}{r^2}
+\left[A'B-C'(D+E)\right]F\epsilon_{(a)\alpha\beta}
\frac{x^\beta}{r}\nonumber\\
&& +\left\{\left[A'B-C'(D+E)\right]D-\frac{1}{r}
\frac{B}{\Delta}\right\}
\left(\delta^\alpha_{(a)}-\frac{x^\alpha x^{(a)}}{r^2}\right)
\Biggr)\psi_{(0)}\;,\nonumber\\
\hat{\mbox{\boldmath $\Sigma$}}_{(a)(b)}^{~~~~~~0\alpha}
&=& \left(2a+3c_2\frac{h'}{r}\right)\Biggl[\frac{BD}{\Delta}
\left(\delta^\alpha_{(a)}\frac{x^{(b)}}{r}
-\delta^\alpha_{(b)}\frac{x^{(a)}}{r}\right) \nonumber\\
&& +\frac{BF}{\Delta}
\left(\epsilon_{(a)\alpha\beta}\frac{x^\beta x^{(b)}}{r^2}
-\epsilon_{(b)\alpha\beta}\frac{x^\beta x^{(a)}}{r^2}\right)\Biggr]
\nonumber \\
&& -\Bigg[2a\Biggl(-\frac{2}{r}C\frac{x^\alpha x^{(a)}}{r^2}
+\left[A'B-C'(D+E)\right]F
\epsilon_{(a)\alpha\beta}\frac{x^\beta}{r}\nonumber\\
&& +\left\{\left[A'B-C'(D+E)\right]D
-\frac{1}{r}\frac{B}{\Delta}\right\}
\left(\delta^\alpha_{(a)}-\frac{x^\alpha x^{(a)}}{r^2}\right)
\Biggr)\psi_{(b)} \nonumber\\
&& -((a)\rightleftarrows (b))\Biggr]\;.
\end{eqnarray}

\subsubsection{Energy-Momentum}
\label{sec:em-choice1-pbar}
By using Eq.~(\ref{eq:super-F2-pbar}) and the expression 
\begin{equation}
\label{eq:em-F1-pbar}
\hat{M}_k=\int\hat{\mbox{\boldmath $F$}}_k^{~0\alpha}
r^2n_\alpha d\Omega 
\end{equation}
which follows from Eqs.~(\ref{eq:M-def-pbar}) and
(\ref{eq:identity-pbar}), we can obtain
\begin{eqnarray}
\label{eq:em-choice1.2-pbar}
\hat{M}_{(0)}&=&16\pi a\lim_{r\to\infty}r\left(
A-\frac{D}{\Delta}\right)\;,\nonumber \\
\hat{M}_{(a)} &=& 0\;.
\end{eqnarray}
We have the relation~\cite{Kawai02} $\hat{M}_k=e^{(0)\mu}_{~~~~k}M_\mu
$ for generic space-times satisfying suitable asymptotic condition.
Here, $M_{\mu }$ is the energy-momentum vector of the total system,
which agrees\footnote{In view of this agreement, we use the same
  symbol $M_{\mu }$ for these two energy-momenta.} with the canonical
energy-momentum in PGT given by Eq.~(\ref{eq:em-psi}).  In view of
this, we require the relation
\begin{equation}
\label{eq:em-p-pbar}
\hat{M}_{k}=-e^{(0)\mu}_{~~~~k}Mc^{2}\delta_{\mu}^{~0}\;,
\end{equation}
following the requirement (\ref{eq:momentum}) in PGT\@.
This is equivalent to
\begin{equation}
\label{eq:AD-condition-pbar}
\lim_{r\to\infty}r\left(\frac{D}{D_{\infty}}
- \frac{A}{A_{\infty}}\right)
= \frac{r_0}{2}\;.
\end{equation}
If we express $A/A_{\infty}$ and $D/D_{\infty}$ as
\begin{equation}
\label{eq:AD-expression-pbar}
\frac{A}{A_{\infty }}=1-\frac{h}{2}+U\;, \; \; 
\frac{D}{D_{\infty }}=1+V\;,
\end{equation}
then the relation $\lim_{r\to\infty}U=0=\lim_{r\to\infty}V$ follows. 
Equation (\ref{eq:AD-condition-pbar}) is equivalent to the condition
\begin{equation}
\label{eq:UV-condition}
\lim_{r\to\infty}r(U-V)=0\;.
\end{equation}

\subsubsection{Angular Momentum}
\label{sec:angular-choice1-pbar}

As is known from Ref.~\citen{Kawai02}, the angular momentum in \= PGT
depends on the asymptotic behavior of the field
$\psi$,\footnote{\label{page:footnote2}We choose this field to be
  regular everywhere in the finite region of the space-time.  This is
  possible, because the field $\psi$ can be chosen arbitrarily as has
  been shown in Ref.~\citen{Kawai01}.} and the following
\begin{eqnarray}
\label{eq:asmpt-psi}
\psi^k &=& e^{(0)k}_{~~~~\mu}x^\mu+\psi^{(0)k}+O(1/r^\beta)\;,\\
\psi^k_{~,\mu} &=& e^{(0)k}_{~~~~\mu}+O(1/r^{1+\beta})\;,
\end{eqnarray}
is quite natural, which we shall assume in this paper. Here,
$\psi^{(0)k}$ and $\beta $ are a constant and a positive constant,
respectively, and $O(1/r^{n})$ with positive $n$ denotes a term for
which $r^{n}O(1/r^{n})$ remains finite for $r \rightarrow \infty $; a
term $O(1/r^{n})$ may of course also be zero.

The angular momentum $\hat{S}_{kl}$ is expressed as 
\begin{equation}
\label{eq:angular-Sigma1-pbar}
\hat{S}_{kl}=\int\hat{\mbox{\boldmath $\Sigma$}}_{kl}^{~~0\alpha}
r^2n_\alpha d\Omega \;,
\end{equation}
which follows from Eqs.~(\ref{eq:S-def-pbar}) and 
(\ref{eq:identity-pbar2}). From Eqs.~(\ref{eq:super-Sigma2-pbar}), 
(\ref{eq:em-p-pbar}) and (\ref{eq:angular-Sigma1-pbar}), 
we obtain the following:
\begin{eqnarray}
\label{eq:S_(0)(a)}
\hat{S}_{(0)(a)}&=&-\hat{M}_{(0)}{\psi^{(0)}}_{(a)}
=2{\psi^{(0)}}_{[(0)}{\hat{M}}_{(a)]}\;,\\
\label{eq:S_(a)(b)}
\hat{S}_{(a)(b)}&=&0=2{\psi^{(0)}}_{[(a)}{\hat{M}}_{(b)]}\;,
\end{eqnarray}
when the conditions (\ref{eq:UV-condition}) and 
\begin{equation}
\label{eq:U-condition}
U=O(1/r^{2})
\end{equation}
are both satisfied. Equations (\ref{eq:S_(0)(a)}) and 
(\ref{eq:S_(a)(b)}) show that $\hat{S}_{kl}$ agrees with the orbital 
angular momentum~\cite{Kawai02} around the origin of the internal 
space ${\mbox{\bf R}}^{4}$. These equations agree with Eq.~(5$\cdot $8) 
of Ref.~\citen{Kawai02}, because the first term in the 
r.h.s.~of this equation vanishes for the Schwarzschild metric.

\subsubsection{Canonical Energy-Momentum 
and \lq\lq Extended Orbital Angular Momentum"}
\label{sec:canonical-choice1-pbar}

The generator $\hat{M}^{c}_{~\mu }$ of coordinate translations and the
generator $\hat{L}_{\mu }^{~\nu }$ of $GL(4,\mbox{\bf R})$ coordinate
transformations are the canonical energy-momentum and\footnote{Note
  that the anti-symmetric part $\hat{L}_{[\mu
    \nu]}\stackrel{\mbox{\scriptsize def}}{=} \hat{L}_{[\mu
    }^{~~\lambda }\eta_{\lambda \nu]}$ is the orbital angular
  momentum.} the \lq\lq extended orbital angular momentum",
respectively. They have the expressions
\begin{eqnarray}
\label{eq:canonical-M-choice1-pbar}
\hat{M}^c_{~\mu} &\stackrel{\mbox{\scriptsize def}}{=}&
\int_{\sigma }{}^{\rm tot}
\hat{\widetilde{\mbox{\boldmath $T$}}}_\mu^{~\raisebox{-.6ex}%
{\scriptsize $\nu$}}
d\sigma_\nu\;,\\
\label{eq:canonical-L-choice1-pbar}
\hat{L}_\mu^{~\nu} &\stackrel{\mbox{\scriptsize def}}{=}&
\int_\sigma\hat{\mbox{\boldmath $M$}}_\mu^{~\nu\lambda}
d\sigma_{\lambda }\;,
\end{eqnarray}
where we have defined\footnote{The energy-momentum density of the
  gravitational field is defined by the r.h.s.~of
  Eq.~(\ref{eq:canonical-TM-choice1-pbar}) with $\hat{\mbox{\boldmath
      $L$}}$ being replaced with $\hat{\mbox{\boldmath $L$}}_{G}$.
  This density does not give vanishing energy-momentum when it is
  integrated over a space-like surface.}
\begin{eqnarray}
\label{eq:canonical-TM-choice1-pbar}
{}^{\rm tot}
\hat{\widetilde{\mbox{\boldmath $T$}}}_\mu^{~\raisebox{-.6ex}%
{\scriptsize $\nu$}}
&\stackrel{\mbox{\scriptsize def}}{=}&
\hat{\mbox{\boldmath $L$}}\delta_\mu^{~\nu}
-\left(\hat{\mbox{\boldmath $F$}}_k^{~\lambda\nu}A^k_{~\lambda,\mu}
+\hat{\mbox{\boldmath $F$}}_{kl}^{~~\lambda\nu}A^{kl}_{~~\lambda,\mu}
+\frac{\partial\hat{\mbox{\boldmath $L$}}}{\partial\psi^k_{~,\nu}}
\psi^k_{~,\mu}+\frac{\partial\hat{\mbox{\boldmath $L$}}}
{\partial\varphi_{,\nu}}\varphi_{,\mu}\right),\qquad\\
\hat{\mbox{\boldmath $M$}}_\mu^{~\nu\lambda}
&\stackrel{\mbox{\scriptsize def}}{=}&
-2\left(
x^\nu~{}^{\rm tot}
\hat{\widetilde{\mbox{\boldmath $T$}}}_\mu^{~\raisebox{-.6ex}%
{\scriptsize $\lambda $}}
-\hat{\mbox{\boldmath $\Psi$}}_\mu^{~\nu\lambda}\right)
\end{eqnarray}
with
\begin{equation}
\label{eq:BoldPsi-choice1-pbar}
\hat{\mbox{\boldmath $\Psi$}}_\mu^{~\nu\lambda}
\stackrel{\mbox{\scriptsize def}}{=}
\hat{\mbox{\boldmath $F$}}_k^{~\nu\lambda}A^k_{~\mu}
+\hat{\mbox{\boldmath $F$}}_{kl}^{~~\nu\lambda}A^{kl}_{~~\mu}\;.
\end{equation}
From Eq.~(\ref{eq:canonical-M-choice1-pbar}) and the identity
\begin{equation}
\label{eq:identity-canonical-choice1-pbar}
-\frac{\delta\hat{\mbox{\boldmath $L$}}}{\delta A^k_{~\nu}}A^k_{~\mu}
-\frac{\delta\hat{\mbox{\boldmath $L$}}}
{\delta A^{kl}_{~~\nu}}A^{kl}_{~~\mu}+
{}^{\rm tot}
\hat{\widetilde{\mbox{\boldmath $T$}}}_\mu^{~\raisebox{-.6ex}%
{\scriptsize $\nu $}}\equiv
\partial_\lambda\hat{\mbox{\boldmath $\Psi$}}_\mu^{~\nu\lambda}\;,
\end{equation}
the expression 
\begin{equation}
\label{eq:canonical-M2-choice1-pbar}
\hat{M}^c_{~\mu}
=\int\hat{\mbox{\boldmath $\Psi$}}_\mu^{~0\alpha}r^2n_\alpha d\Omega \;,
\end{equation}
is obtained. By evaluating the r.h.s.~of this, we obtain
\begin{equation}
\label{eq:canonical-M3-choice1-pbar}
\hat{M}^c_{~\mu}=0\;,
\end{equation}
when the condition
\begin{equation}
\label{eq:rU-condition}
\lim_{r\to \infty}rU=0\;.
\end{equation}
is satisfied.
For $\hat{L}_{\mu }^{~\nu }$, we have 
\begin{equation}
\label{eq:orbital-L2-choice1-pbar}
\hat{L}_\mu^{~\nu}
=-2\int x^\nu\hat{\mbox{\boldmath $\Psi$}}_\mu^{~0\alpha}
n_\alpha d\Omega \;,
\end{equation}
which follows from Eqs.~(\ref{eq:canonical-L-choice1-pbar}) and
(\ref{eq:identity-canonical-choice1-pbar}). By evaluating the
r.h.s.~of this, we obtain
\begin{equation}
\label{eq:anti-symmetric-orbital-L3-pbar}
\hat{L}_{0}^{~0}=0\;,\; \; 
\hat{L}_{\mu }^{~\nu }=0\;, \; \; \mu \neq \nu \;,
\end{equation}
when the conditions (\ref{eq:UV-condition}) and (\ref{eq:rU-condition})
are both satisfied.
Also, we have  
\begin{equation}
\label{eq:orbital-L3-choice1-pbar}
\hat{L}_{1}^{~1}=\hat{L}_{2}^{~2}=\hat{L}_{3}^{~3}=0\;, 
\end{equation}
if the conditions (\ref{eq:rU-condition}) and 
\begin{equation}
\label{eq:U'-condition}
U'=O(1/r^{1+\gamma })
\end{equation}
with $\gamma $ being a positive constant are both satisfied.  It is
also known that the orbital angular momentum $\hat{L}_{[\mu \nu ]}$
vanishes if the condition (\ref{eq:U-condition}) is satisfied.

\subsection{The case when $\{\psi^k,e^k_{~\mu},A^{kl}_{~~\mu},
\varphi \}$ is employed as the set of independent field 
variables}
\label{sec:choice2-pbar}

Let us denote the Lagrangians $\mbox{\boldmath $L$}$ and 
$\mbox{\boldmath $L$}_{G}$ expressed as 
functions of $\psi^k,e^k_{~\mu},A^{kl}_{~~\mu},\varphi $ and of 
their derivatives by $\check{\mbox{\boldmath $L$}}$ and 
$\check{\mbox{\boldmath $L$}}_{G}$, respectively.

For the dynamical energy-momentum $\check{M}_k$, we 
have~\cite{Kawai-Saitoh}
\begin{equation}
\label{eq:M-choice2-pbar}
\check{M}_k \stackrel{\mbox{\scriptsize def}}{=}
\int_\sigma{}^{\rm tot}\check{\mbox{\boldmath $T$}}_k^{~\mu}
d\sigma_\mu\equiv0 
\end{equation}  
with
\begin{equation}
\label{eq:MDchoice2-pbar}
{}^{\rm tot}\check{\mbox{\boldmath $T$}}_k^{~\mu}
\stackrel{\mbox{\scriptsize def}}{=}
\frac{\partial\check{\mbox{\boldmath $L$}}}{\partial\psi^k_{~,\mu}}
+i\frac{\partial\check{\mbox{\boldmath $L$}}}{\partial\varphi_{,\mu}}
P_k\varphi \equiv 0\;.
\end{equation}
The generator $\check{S}_{kl}$ of internal Lorentz transformations is 
expressed as  
\begin{equation}
\label{eq:S-choice2-pbar}
\check{S}_{kl} \stackrel{\mbox{\scriptsize def}}{=}
\int_\sigma{}^{\rm tot}\check{\mbox{\boldmath $S$}}_{kl}^{~~\mu}
d\sigma_\mu 
\end{equation}
with
\begin{equation}
\label{eq:S2-choice2-pbar}
{}^{\rm tot}\check{\mbox{\boldmath $S$}}_{kl}^{~~\mu}
\stackrel{\mbox{\scriptsize def}}{=}
-2\frac{\partial\check{\mbox{\boldmath $L$}}}
{\partial\psi^{[k}_{~~,\mu}}
\psi_{l]}-2\frac{\partial\check{\mbox{\boldmath $L$}}}
{\partial e^{[k}_{~~\nu,\mu}}
e_{l]\nu}-4\frac{\partial\check{\mbox{\boldmath $L$}}}
{\partial A^{[k}_{~~m\nu,\mu}}A_{l]m\nu}
-i\frac{\partial\check{\mbox{\boldmath $L$}}}
{\partial \varphi_{,\mu}}M_{kl}\varphi \;.
\end{equation}
Also, we have the identity 
\begin{equation}
\label{eq:identity-choice2-pbar}
\frac{\delta\check{\mbox{\boldmath $L$}}}{\delta A^{kl}_{~~\mu}}
+\frac{1}{2}{}^{\rm tot}\check{\mbox{\boldmath $S$}}_{kl}^{~~\mu}
\equiv\frac{1}{2}
\partial_{\nu }\check{\mbox{\boldmath $\Sigma$}}_{kl}^{~~\mu\nu}
\end{equation}
with
\begin{equation}
\label{eq:super-Sigma-choice2-pbar}
\check{\mbox{\boldmath $\Sigma$}}_{kl}^{~~\mu\nu}
\stackrel{\mbox{\scriptsize def}}{=}
-2\frac{\partial\check{\mbox{\boldmath $L$}}}
{\partial A^{kl}_{~~\mu,\nu}}
+4a\left(\sqrt{-g}e^{[\mu}_{~~k}e^{\nu]}_{~~l}
-e^{(0)[\mu}_{~~~~~k}e^{(0)\nu]}_{~~~~~l}\right)\;.
\end{equation}

Equation (\ref{eq:S-choice2-pbar}) can be rewritten as
\begin{equation}
\label{eq:S3-choice2-pbar}
\check{S}_{kl}=\int\check{\mbox{\boldmath $\Sigma$}}_{kl}^{~~0\alpha}
r^2n_\alpha d\Omega \;,
\end{equation}
with the aid of Eq.~(\ref{eq:identity-choice2-pbar}). The integrand in
Eq.~(\ref{eq:S3-choice2-pbar}) agrees with that in
Eq.~(\ref{eq:spin-angular-p}), and it follows that
\begin{equation}
\label{eq:S-p-pbar}
\check{S}_{kl}=S_{kl}=0 \;,
\end{equation}
which holds for any $A,B,C,D,E$ and $F$ satisfying 
the conditions (\ref{eq:AF-condition}) and (\ref{eq:AD-condition}). 

The canonical energy-momentum $\check{M}^c_{~\mu}$ is expressed as  
\begin{equation}
\label{eq:canonical-M-choice2-pbar}
\check{M}^c_{~\mu} \stackrel{\mbox{\scriptsize def}}{=}
\int_\sigma
{}^{\rm tot}
\check{\widetilde{\mbox{\boldmath $T$}}}_\mu^{~\raisebox{-.6ex}%
{\scriptsize $\nu$}}
d\sigma_\nu 
\end{equation}
with\footnote{The energy-momentum density of the gravitational field
  is defined by the r.h.s.~of Eq.~(\ref{eq:T-choice2-pbar}) with
  $\check{\mbox{\boldmath $L$}}$ being replaced with
  $\check{\mbox{\boldmath $L$}}_{G}$. This density agrees with that of
  PGT, if the \lq \lq intrinsic" energy-momentum $P_{k}$ of the field
  $\varphi $ is vanishing.}
\begin{equation}
\label{eq:T-choice2-pbar}
{}^{\rm tot}
\check{\widetilde{\mbox{\boldmath $T$}}}_\mu^{~\raisebox{-.6ex}%
{\scriptsize $\nu$}}
\stackrel{\mbox{\scriptsize def}}{=}
\check{\mbox{\boldmath $L$}}\delta_\mu^{~\nu}
-\left(\frac{\partial\check{\mbox{\boldmath $L$}}}
{\partial\psi^k_{~,\nu}}\psi^k_{~,\mu}
+\frac{\partial\check{\mbox{\boldmath $L$}}}
{\partial e^k_{~\lambda,\nu}}e^k_{~\lambda,\mu}
+\frac{\partial\check{\mbox{\boldmath $L$}}}
{\partial A^{kl}_{~~\lambda,\nu}}
A^{kl}_{~~\lambda,\mu}+\frac{\partial\check{\mbox{\boldmath $L$}}}
{\partial\varphi_{,\nu}}\varphi_{,\mu}\right)\;.
\end{equation}
There is the identity
\begin{equation}
\label{eq:identity-canonical-choice3-pbar}
-\frac{\delta\check{\mbox{\boldmath $L$}}}{\delta e^k_{~\nu}}e^k_{~\mu}
-\frac{\delta\check{\mbox{\boldmath $L$}}}{\delta A^{kl}_{~~\nu}}
A^{kl}_{~~\mu}
+{}^{\rm tot}
\check{\widetilde{\mbox{\boldmath $T$}}}_\mu^{~\raisebox{-.6ex}%
{\scriptsize $\nu$}}\equiv 
\partial_\lambda\check{\mbox{\boldmath $\Psi$}}_\mu^{~\nu\lambda}\;,
\end{equation}
where we have defined 
\begin{equation}
\label{eq:super-Psi-choice2-pbar}
\check{\mbox{\boldmath $\Psi $}}_\mu^{~\nu\lambda}
\stackrel{\mbox{\scriptsize def}}{=}
\frac{\partial\check{\mbox{\boldmath $L$}}}
{\partial e^k_{~\nu,\lambda}}e^k_{~\mu}
+\frac{\partial\check{\mbox{\boldmath $L$}}}
{\partial A^{kl}_{~~\nu,\lambda}}
A^{kl}_{~~\mu}\;.
\end{equation}
From Eqs.~(\ref{eq:M-def-p}), (\ref{eq:super-po1-p}), 
(\ref{eq:identity-p}), (\ref{eq:canonical-M-choice2-pbar}),
(\ref{eq:identity-canonical-choice3-pbar}) and 
(\ref{eq:super-Psi-choice2-pbar}), we can show that 
\begin{equation}
\label{eq:canonical-em-p-pbar}
\check{M}^c_{~\mu}=M_\mu\;.
\end{equation}
When the requirement (\ref{eq:momentum}) is imposed, the functions 
$A,B,C,D,E$ and $F$ agree with as those in the case of PGT.

The \lq \lq extended orbital angular momentum", which is the generator
of $GL(4,\mbox{\bf R})$ coordinate transformations, has the expression 
\begin{equation}
\label{eq:orbital-ang-choice2-pbar}
\check{L}_\mu^{~\nu} \stackrel{\mbox{\scriptsize def}}{=}
\int_\sigma\check{\mbox{\boldmath $M$}}_\mu^{~\nu\lambda}
d\sigma_\lambda 
\end{equation}
with
\begin{equation}
\label{eq:orbital-M-choice2-pbar}
\check{\mbox{\boldmath $M$}}_\mu^{~\nu\lambda}
\stackrel{\mbox{\scriptsize def}}{=}
-2\left(x^{\nu}~{}^{\rm tot}
\check{\widetilde{\mbox{\boldmath $T$}}}_\mu^{~\raisebox{-.6ex}%
{\scriptsize $\lambda$}}
-\check{\mbox{\boldmath $\Psi $}}_\mu^{~\nu\lambda}\right)\;.
\end{equation}
The integral in Eq.~(\ref{eq:orbital-ang-choice2-pbar}) is evaluated 
to give 
\begin{eqnarray}
\label{eq:orbital-ang2-choice2-pbar}
\check{L}_{[\mu \nu]}&\stackrel{\mbox{\scriptsize def}}{=}&
\check{L}_{[\mu }^{~~\lambda }\eta_{\lambda \nu]}
=0\;,\nonumber \\
\check{L}_\mu^{~0}&=&-2x^0M_\mu\;,
\quad \check{L}_0^{~\alpha}=0\;,\nonumber\\
\check{L}_\alpha^{~\beta}&=&\frac{16\pi}{3}\delta_\alpha^{~\beta}
\lim_{r\rightarrow \infty }
\left\{\frac{2ar^{2}BD}{\Delta} +3c_{2}r_{0}[A'B-C'(D+E)]\right\}\;,
\end{eqnarray}
the first of which implies that the orbital angular momentum is vanishing. 
We have 
\begin{equation}
\label{eq:orbital-diag}
\check{L}_{1}^{~1}=\check{L}_{2}^{~2}=\check{L}_{3}^{~3}=0\;,
\end{equation}
if 
\begin{equation}
\label{eq:Y-condition}
P=-\frac{1}{2^{3}}h^{2}-\frac{1}{2^{4}}h^{3}-\frac{5}{2^{7}}h^{4}+Z
\end{equation}
with $Z$ satisfying $Z=O(1/r^{6})$ and $\lim_{r\to \infty }r^{3}Z'=0$.

\section{Summary and comments}
\label{sec:summary}

The results obtained in \S\S 3$\sim $6 can be summarized as follows:
\begin{enumerate}
\item Both in PGT and in \= PGT, the Schwarzschild space-time
  expressed in terms of the Schwarzschild coordinates is obtainable as
  a torsionless exact solution of gravitational field equations with a
  spinless point-like source located at the origin, if and only if the
  condition (\ref{eq:parameter-condition}) is satisfied. This and the
  fact mentioned in the footnote on page \pageref{page:footnote} show
  that the Lagrangian $L_{R}$ with the condition
  (\ref{eq:parameter-condition}) is favorable in various respects.
\item Spherically symmetric vierbeins (\ref{eq:vierbein}) have been
  considered by choosing the gauge of internal Lorentz group in a way
  such that the function $A$ is independent of $x^{0}$.
\item For PGT, the equality of the active gravitational mass and the
  inertial mass is satisfied, if and only if the condition
  (\ref{eq:AD-condition-p}), which is equivalent to
  Eq.~(\ref{eq:YP-condition}), is satisfied. Also, the spin angular
  momentum $S_{kl}$ and orbital angular momentum $L^{\mu \nu }$ both
  vanish for any $A,B,C,D,E$ and $F$ satisfying the conditions
  (\ref{eq:AF-condition}) and (\ref{eq:AD-condition}).
\item For \=PGT, generators depend on the choice of the set of
  independent field variables.
\begin{description}
\item[(A)] The case when $\{\psi^{k}, A^{k}_{~\mu }, A^{kl}_{~~\mu
    },\varphi \}$ is employed as the set of independent
  field variables. \newline
  For this case, dynamical energy-momentum $\hat{M}_{k}$, which is the
  generator of internal translations, has the expression
  (\ref{eq:em-p-pbar}),\footnote{Remind the paragraph just below
    Eq.~(\ref{eq:em-choice1.2-pbar}) and note that the relation
    $M_{\mu }=-Mc^{2}\delta_{\mu }^{~0}$ gives the equality of the
    active gravitational mass and the inertial mass.} if the condition
  (\ref{eq:AD-condition-pbar}) (or equivalently the condition
  (\ref{eq:UV-condition})) is satisfied. For the total angular
  momentum $\hat{S}_{kl}$, we have
  $\hat{S}_{kl}=2{\psi^{(0)}}_{[k}\hat{M}_{l]}$, if the conditions
  (\ref{eq:UV-condition}) and (\ref{eq:U-condition}) are both
  satisfied.  This shows that the total angular momentum of this
  space-time is only the orbital angular momentum around the origin of
  the internal space ${\bf R}^{4}$, which is quite reasonable because
  this space-time is a static spherically symmetric one with a static
  spinless point-like source. The canonical energy-momentum
  $\hat{M}^{c}_{~\mu}$ vanishes, if the condition
  (\ref{eq:rU-condition}) is satisfied. The \lq \lq extended orbital
  angular momentum" $\hat{L}_{\mu }^{~\nu}$ vanishes, if the
  conditions (\ref{eq:rU-condition}) and (\ref{eq:U'-condition}) are
  both satisfied. Thus, reasonable energy-momentum and angular
  momentum are obtained as the generators of the {\em internal\/}
  Poincar\'e transformations and the generators of general affine {\em
    coordinate\/} transformations vanish, if the conditions
  (\ref{eq:UV-condition}), (\ref{eq:U-condition}),
  (\ref{eq:rU-condition}) and (\ref{eq:U'-condition}) are all
  satisfied.
\item[(B)] The case when $\{\psi^{k}, e^{k}_{~\mu },
  A^{kl}_{~~\mu},\varphi \}$ is employed as the set of independent
  field variables.\newline
  For this case, the dynamical energy-momentum $\check{M}_{k}$
  identically vanish,~\cite{Kawai-Saitoh} and also the spin angular
  momentum $\check{S}_{kl}$ and the orbital angular momentum
  $\check{L}_{[\mu \nu]}$ both vanish for any $A,B,C,D,E$ and $F$
  satisfying the conditions (\ref{eq:AF-condition}) and
  (\ref{eq:AD-condition}).  The canonical energy-momentum
  $\check{M}^{c}_{~\mu }$ agrees with the energy-momentum $M_{\mu }$
  in PGT, and the active gravitational mass is equal to the inertial
  mass, if the condition (\ref{eq:AD-condition-p}) is satisfied. The
  \lq \lq extended orbital angular momentum" $\check{L}_{\mu }^{~\nu
    }$ vanishes, if the condition (\ref{eq:Y-condition}) is satisfied.
\end{description}
\end{enumerate}

Finally, we would like to add several comments:
\begin{description}
\item[{[1]}] In general relativity, the same expression as
  (\ref{eq:emd}) has been obtained\cite{Kawai-Sakane} for the
  energy-momentum density of the source of the Schwarzschild
  space-time.  Thus, the gravitational field of this space-time can be
  interpreted to be produced by a point-like particle with mass $M$
  located at $\mbox{\boldmath{$x$}}=\mbox{\boldmath{$0$}}$.  The
  regularization scheme employed is crucial in giving this
  result.\cite{Kawai-Sakane} In the present paper, we have developed
  our discussion on the premise that this interpretation and the
  regularization scheme are both applicable also to PGT and to \=PGT.
\item[{[2]}] For the Schwarzschild space-time, the Lorentz gauge
  potentials $A^{kl}_{~~\mu }$ agrees with the Ricci rotation
  coefficients $\Delta^{kl}_{~~\mu }$, and the results given in \S 5.
  and in \S 6.  show that reasonable energy-momenta and angular
  momenta are obtainable, even if the condition~\cite{HS02,Kawai02}
\begin{equation}
\label{eq:Ricci-A}
\Delta^{kl}_{~~\mu }=O\left(\frac{1}{r^{1+\alpha }}\right)\;, \; \; 
\Delta^{kl}_{~~\mu,(m)}=O\left(\frac{1}{r^{2+\alpha }}\right)\;,\; \; 
m=1,2, \; \; \alpha >0\;,  
\end{equation}
or the weaker condition~\cite{HS02,Kawai02}
\begin{equation}
\label{eq:Ricci-B}
\Delta^{kl}_{~~\mu }=\frac{M^{kl}_{~~\mu }}{r}+L^{kl}_{~~\mu }
\end{equation}
with $M^{kl}_{~~\mu }$ being a constant and 
\begin{equation}
\label{eq:L}
L^{kl}_{~~\mu }=O\left(\frac{1}{r^{2}}\right)\;, \; \; 
L^{kl}_{~~\mu ,(m)}=O\left(\frac{1}{r^{3}}\right)\;, \; \; m=1,2\;,
\end{equation}
are not satisfied. Here, $\Delta^{kl}_{~~\mu,(m)}$, for example, 
denotes the $m$-th order partial derivative of 
$\Delta^{kl}_{~~\mu}$ with respect to $x^{\lambda }$.

\item[{[3]}] As we have seen in the summary, the requirement that the
  active gravitational mass is equal to the inertial mass restricts
  the behavior of the vierbeins at spatial infinity. The physical
  meaning of vierbeins violating this equality is not yet clear. In
  this connection, it is worth mentioning the following:~(1) There is
  a similar situation also in new
  general relativity.~\cite{Shirafuji-Nashed-Hayashi,%
Shirafuji-Nashed-Kobayashi} (2) Also in the framework of general
  relativity, the equality of the active gravitational mass and the
  inertial gravitational mass is violated for the Schwarzschild metric
  expressed with the use of a certain coordinate
  system.~\cite{Bozhkov-Rodrigues}
\item[{[4]}] Discussions in \S 5.~and in \S 6.~present us with a test
  of definitions of energy-momenta and of angular momenta introduced
  in Ref.~\citen{HS02} and in Ref.~\citen{Kawai02}, and the results
  summarized in 3.~and in 4.~support them.
\end{description}

\appendix
\section{Irreducible components of $T_{klm}$ and of $R_{klmn}$}
\label{appendix:1}
Irreducible components of $T_{klm}$ and of $R_{klmn}$ are the following: 
\begin{eqnarray}
\label{eq:irre-comp-TR}
t_{klm} &\stackrel{\mbox{\scriptsize def}}{=}&
\frac{1}{2}(T_{klm}+T_{lkm})+\frac{1}{6}(\eta_{mk}v_l+\eta_{ml}v_k)
-\frac{1}{3}\eta_{kl}v_m\;,\\
v_k &\stackrel{\mbox{\scriptsize def}}{=}& T^l_{~lk}\;,\\
a_k &\stackrel{\mbox{\scriptsize def}}{=}&
\frac{1}{6}\epsilon_{klmn}T^{lmn}\;, \\
A_{klmn} &\stackrel{\mbox{\scriptsize def}}{=}&
\frac{1}{6}(R_{klmn}+R_{kmnl}+R_{knlm}
+R_{lmkn}+R_{lnmk}+R_{mnkl}) \;,\\
B_{klmn} &\stackrel{\mbox{\scriptsize def}}{=}&
\frac{1}{4}(W_{klmn}+W_{mnkl}-W_{knlm}-W_{lmkn}) \;,\\
C_{klmn} &\stackrel{\mbox{\scriptsize def}}{=}&
\frac{1}{2}(W_{klmn}-W_{mnkl}) \;,\\
E_{kl} &\stackrel{\mbox{\scriptsize def}}{=}&
\frac{1}{2}(R_{kl}-R_{lk}) \;,\\
I_{kl} &\stackrel{\mbox{\scriptsize def}}{=}&
\frac{1}{2}(R_{kl}+R_{lk})-\frac{1}{4}\eta_{kl}R\;,\\
R &\stackrel{\mbox{\scriptsize def}}{=}&
\eta^{kl}R_{kl}
\end{eqnarray}
with
\begin{eqnarray}
\label{eq:irre-comp2}
W_{klmn} &\stackrel{\mbox{\scriptsize def}}{=}&
R_{klmn}-\frac{1}{2}(\eta_{km}R_{ln}+\eta_{ln}R_{km}
-\eta_{kn}R_{lm}-\eta_{lm}R_{kn}) \nonumber \\
&&\quad +\frac{1}{6}(\eta_{km}\eta_{ln}-\eta_{lm}\eta_{kn})R \;,\\
R_{kl} &\stackrel{\mbox{\scriptsize def}}{=}&
\eta^{mn}R_{kmln}\;.
\end{eqnarray}

\section{Vierbeins giving the Schwarzschild metric}
\label{appendix:2}
Vierbeins having the expression (\ref{eq:vierbein}) are fixed by 
the functions $A,B,C,D,E$ and $F$. We write down here an example 
of the set of $A,B,\cdots $ and $F$ which gives the Schwarzschild 
metric: 
\begin{eqnarray}
\label{eq:A}
A&=&\pm \left(1-\frac{h}{2}+Jh^{\omega }\right)\;, \\ 
\label{eq:D}
D&=&\pm \frac{1}{1+K(x^{0})h^{\omega }}\;,\\ 
\label{eq:CF}
C&=&\sqrt{A^{2}-1+h}\;,\; \; F=\sqrt{1-D^{2}}\;, \\
\label{eq:B}
B&=&\pm \frac{C}{1-h}\;, \\
\label{eq:E}
E&=&\pm \frac{A}{1-h}-D
\end{eqnarray}
with the double signs in the expressions for $B$ and $E$ in same 
order.\footnote{Otherwise, the order of the double signs in 
Eqs.~(\ref{eq:A})$\sim $ (\ref{eq:E}) is arbitrary.} Also, $\omega $,
$J$ and $K$ are a real constant, a non-negative 
real constant and a non-negative real valued function of $x^{0}$, 
respectively. 

The function $A$ in the above is independent of $x^{0}$, and we have 
the following:
\begin{eqnarray}
\label{eq:limit-1}
\Delta &=&\pm 1\;, \\
\label{eq:limit-2}
A_{\infty }&=&\pm 1\;, \; \mbox{\rm if and only if}\; J=0\; \; 
\mbox{\rm or}\; \omega >0\;, \\
\label{eq:limit-3}
D_{\infty }&=&\pm 1\;, \; \mbox{\rm if and only if}\; K\equiv 0\; \; 
\mbox{\rm or}\; \omega >0\;, 
\end{eqnarray}
where the double signs in Eq.~(\ref{eq:limit-1}), Eq.~(\ref{eq:limit-2})
and Eq.~(\ref{eq:limit-3}) are in same order as those
in Eqs.~(\ref{eq:B}) and (\ref{eq:E}), Eq.~(\ref{eq:A}) and 
Eq.~(\ref{eq:D}), respectively.
The restrictions imposed on $\omega , J$ and $K$ by the asymptotic 
conditions (\ref{eq:YP-condition}), (\ref{eq:UV-condition}), 
(\ref{eq:U-condition}), (\ref{eq:rU-condition}), 
(\ref{eq:U'-condition}), etc.~are easily known.

\end{document}